\documentclass[prl,twocolumn,showpacs,floatfix]{revtex4}

\usepackage{times,txfonts,graphicx}

\newcommand{\half}{\frac{1}{2}}

\newcommand{\Vdd}{V_{\mathrm{dd}}}
\newcommand{\rr}{{\bf r}}
\newcommand{\rrp}{{\bf r'}}
\newcommand{\sixth}{\frac{1}{6}}
\mathchardef\ogon="012C%
\newcommand{\as}{a\kern-0.22em\lower.40ex\hbox{$_{\ogon}$}}
 
\begin{document}
%\draft

\title{Hydrodynamic excitations of trapped dipolar fermions}
 
\author{Krzysztof G{\'o}ral,$^1$ Miros{\l}aw Brewczyk,$^2$ and 
Kazimierz Rz{\c a}\.zewski$\,^1$}
\affiliation{$^1$Center for Theoretical Physics, Polish Academy of 
Sciences, Aleja Lotnik\'ow 32/46, 02-668 Warsaw, Poland \\ 
$^2$Uniwersytet w Bia{\l}ymstoku, ul. Lipowa 41, 15-424 Bia{\l}ystok, 
Poland}
 
%\date{\today}
%\maketitle

\begin{abstract}
A single-component Fermi gas of polarized dipolar particles in a harmonic 
trap can undergo a mechanical collapse due to the attractive part of the 
dipole-dipole interaction. This phenomenon can be conveniently 
manipulated by the shape of the external trapping potential. We 
investigate the signatures of the instability by studying the spectrum of 
low-lying collective excitations of the system in the hydrodynamic 
regime. To this end, we employ a time-dependent variational method as 
well as exact numerical solutions of the hydrodynamic equations of the 
system.
\end{abstract}
 

\maketitle

%\begin{multicols}{2}

Following a spectacular series of experimental achievements with 
Bose-Einstein condensates in dilute trapped atomic gases (for a review 
see, e.g., \cite{natureAnglin,natureRolston}), dilute gaseous atomic Fermi 
systems have drawn intensified 
interest only very recently. After the first successful realization of 
quantum degeneracy in a potassium gas \cite{fermions}, several other 
groups have cooled fermionic $^6$Li below the Fermi temperature and 
trapped it magnetically or optically 
\cite{fermionsHulet,fermionsENS,OHara}. Very recently, two-species 
Fermi-Bose mixtures have been also achieved in $^{6}$Li--$^{23}$Na 
\cite{fermionsMIT} and in $^{40}$K--$^{87}$Rb \cite{modugno}. Much 
emphasis has been put on the attainment of the transition to a superfluid 
Cooper-paired state \cite{fermionsHolland,fermionsEddy,cirac}.

In a single-component Fermi gas at very low energies s-wave scattering is 
excluded by the statistics. Therefore, such a system behaves as a nearly 
ideal one. However, in this situation other types of interactions, usually 
negligible, come into play. A good candidate is a dipole-dipole force. Its 
effects might be seen in the behavior of atoms with large permanent 
magnetic moments (such as chromium \cite{chromiumPfau,chromiumHarvard}) 
and should be evident 
in the case of heteronuclear molecules possessing permanent electric 
dipoles (an example is ND$_3$, which has been cooled in a Stark 
decelerator down to the millikelvin range \cite{MeijerNature1}). Another 
option is to induce electric dipoles with the help of strong DC fields 
\cite{Marinescu} or by optical admixing of large dipole moments of Rydberg 
states \cite{Santos}.

The ground state of a harmonically trapped polarized single-component 
dipolar Fermi gas in the normal phase has been investigated in 
\cite{fermionsGoral}. An interesting possibility of mechanical instability 
of the system has been investigated in detail. Using the Thomas-Fermi 
theory (together with the Dirac correction) the authors have shown that 
when the critical values of the total particle number $N$ or of the dipole 
moment $d$ are reached, the system collapses (for an analogous discussion 
of a dipolar Bose condensate see \cite{Santos}). It has been shown that 
this behavior is governed by a parameter $N^\sixth\varepsilon$, where

\begin{equation}
\label{eq:scale8b}
\varepsilon=\left(\omega m^3/\hbar^5\right)^{\half} d^2 \,.
\end{equation}

\noindent In Eq.(\ref{eq:scale8b}) $\omega$ is the transverse frequency 
of an axially symmetric harmonic trap (the trap axis being parallel to the 
dipole polarization) and $m$ is the particle mass. The critical value of 
$N^\sixth\varepsilon$ depends on the trap aspect ratio 
$\beta=\omega_z/\omega$ ($\omega_z$ is the axial frequency of the trap). A 
detailed stability diagram is presented in Fig.1 of 
Ref.\cite{fermionsGoral}. 
Here we summarize its main features pertinent to the following discussion. 
The most striking finding is the existence of a characteristic value of 
the aspect ratio $\beta\simeq5.2$ above which (i.e., in sufficiently 
oblate 
traps) the system is always stable, for any number of 
particles and any dipole moment. Away from this region the critical values 
of $N^\sixth\varepsilon$, above which the system is unstable, tend to lie 
between 2 and 10. Generally, the critical $N^\sixth\varepsilon$ is 
smaller for traps elongated in the direction of polarization, i.e., for 
$\beta<1$ (with an exception region for extremely stiff traps with 
$\beta \ll 1$, where the critical value of $N^\sixth\varepsilon$ grows 
again). As regards the experimental prospects of observing the predicted 
instability, one can say that it will definitely not affect weak magnetic 
dipoles \cite{chromiumPfau,chromiumHarvard}. However, polar molecules 
and strong electrically induced atomic dipoles may very well be located in 
the vicinity of the unstable region.

At this point, a discussion of p-wave induced collapse in a 
single-component Fermi gas should also be mentioned \cite{Roth}. Another 
interesting aspect of the physics of trapped dipolar fermions, an issue of 
superfluid pairing induced by the dipole-dipole interactions, has also 
been addressed recently in Refs.\cite{dipolesBCS,physica}.

The issue of collective excitations of degenerate Fermi gases is important
and its studies have the same motivations as in the case of Bose
condensates -- the spectrum of collective excitations determines the
system dynamics in the low-energy regime. This way, the analysis of the
excitations brings yet another important piece of information about
quantum degenerate fermionic gases. A few works have been concerned with
the problem. Amoruso {\it et al.} \cite{Amoruso} as well as Bruun and
Clark \cite{fermionsClark}  have investigated hydrodynamic excitations in
non-interacting Fermi gases, mainly in spherically symmetric traps. Their
work has been extended to the case of anisotropic trapping potentials by
Csord\'{a}s and Graham \cite{Csordas}. Some experimental work should be
mentioned in the context of collective excitations -- DeMarco and Jin have 
investigated spin excitations in a fermionic gas of atoms \cite{SpinExc}.
 
In the case of degenerate bosons, the spectrum of the collective
excitations strongly depends on the particle interactions. In the 
mean-field picture, this amounts to the dependence on both the number of 
particles and the scattering length. As single-component degenerate 
fermions at low energies are virtually non-interacting, the problem of 
interactions is less important and appears in two contexts only: either
one deals with a two-component gas (where inter-component scattering
takes place) \cite{Vichi} or, in the single-component case, one takes
into account dipole-dipole forces. In this paper we study the latter case.
 
In the hydrodynamic regime, i.e., when collisions ensure local 
equilibrium,
the dynamics of a Fermi gas is governed by the hydrodynamic equations for
the fields of velocity ${\bf v}(\rr,t)$ and density $n(\rr,t)$:
\begin{eqnarray}
&&\frac{\partial n(\rr,t)}{\partial t} = 
- \nabla \cdot \left[n(\rr,t) {\bf v}(\rr,t)\right]  \nonumber \\
&&\frac{\partial {\bf v}(\rr,t)}{\partial t} =
- \nabla \left[ \frac{\hbar^2}{2 m^2} \left[6\pi^2 n(\rr,t) \right]^{2/3}
+ \frac{1}{2} {\bf v}^2(\rr,t)  \right. \nonumber \\
&&\left. +\frac{1}{2}\omega^2 \left[ x^2+y^2+(\beta z)^2 \right]
+\frac{1}{m} \int d\rrp \, n(\rr,t) \, \Vdd(\rr-\rrp) \right] \, ,
\nonumber \\
\label{hydrodynamic}
\end{eqnarray}
where $\Vdd(\rr)$ is the dipole-dipole potential
\begin{equation}
\label{eq:Vdd1}
\Vdd({\rr})=
\frac{d^2}{r^3}-3\frac{({\bf d}\cdot{\rr})^2}{r^5} \, .
%-\frac{8\pi}{3}d^2\delta({\rr})\, .
\end{equation}

In the second of Eqs.(\ref{hydrodynamic}) one can recognize various 
components of the Thomas-Fermi formulation for trapped dipolar fermions 
\cite{fermionsGoral}: the first term on the right-hand side is the quantum 
pressure (microscopic kinetic energy) resulting from Fermi statistics; the 
next term describes the macroscopic part of the kinetic energy; the first 
term in the last line is the harmonic trapping potential and the last term 
in this line is the dipole-dipole interaction energy.

In order to study low-lying collective
excitations, in the following we propose to merge the variational approach 
(used to find the ground state of a dipolar Fermi gas in 
Ref.\cite{fermionsGoral}) with the time-dependent variational method used 
to study bosonic oscillations (see, e.g., Refs. 
\cite{Perez,Yi2,ExcBosons}). A related approach has been developed by 
Zaremba and Tso \cite{Zaremba} to model a parabolically confined electron 
gas.

First, we introduce the Lagrangian density

\begin{eqnarray}
&&{\cal L}(\rr,t)=-\hbar \, n(\rr,t)\frac{\partial \chi(\rr,t)}{\partial
t} -\frac{\hbar^2}{m}\frac{1}{20 \pi^2} \left[6\pi^2 n(\rr,t)
\right]^{5/3} \nonumber \\
&&-\frac{\hbar^2}{2m} n(\rr,t) (\nabla \chi(\rr,t))^2
-\frac{1}{2} m\omega^2 \left[ x^2+y^2+(\beta z)^2 \right] \, n(\rr,t) 
\nonumber \\
&&-\frac{1}{2} \int d\rrp \, n(\rr,t) \, \Vdd(\rr-\rrp) \, n(\rrp,t) \, 
, 
\label{Lfermions}
\end{eqnarray}
 
\noindent where ${\bf v}(\rr,t)=(\hbar/2m) \lim_{\vec{r}_1 \rightarrow
\vec{r}_2} (\vec{\nabla}_1 - \vec{\nabla}_2) 
\chi(\vec{r}_1,\vec{r}_2,t)$ and $n(\rr,t)$ is the spatial one-particle 
density. $\chi$ is usually termed the potential of the velocity field. 
% as defined by the one-particle Wigner function $\nu({\rr},{\bf p})$ 
%\begin{equation}
%n({\rr},t)= n^{(1)}({\rr};{\rr},t)
%=\int\frac{ \D{\bf p} }{(2\pi\hbar)^3} \, \nu({\rr},{\bf p},t)\,.
%\label{eq:Wig2}
%\end{equation}
The Lagrangian density ${\cal L}(\rr,t)$ is chosen such that the 
Lagrange equations for the conjugate variables $n$ and $\chi$, following 
from the Lagrangian $L=\int d\rr \, {\cal L}(\rr,t)$, reproduce the 
hydrodynamic equations (\ref{hydrodynamic}) for the Fermi gas.
 
From this point we proceed in a close analogy to the procedure employed 
for bosons in Ref.\cite{ExcBosons}. We introduce a Gaussian variational 
ansatz for the density $n(\rr,t)$ :

\begin{equation}
\label{ansatz}
n(x,y,z,t)=\frac{N}{\pi^{3/2}w_{x}(t)w_{y}(t)w_{z}(t)}
\prod_{\eta=x,y,z}
%\exp\left(-\left[\eta-\eta_{0}(t)\right]^2/w^2_{\eta}(t) \right)
%e^{\frac{-\left[\eta-\eta_{0}(t)\right]^2}{w^2_{\eta}(t)}}
e^{-\left[\eta-\eta_{0}(t)\right]^2/w^2_{\eta}(t)} \; ,
\end{equation}

\noindent where $w_{\eta}(t)$ are time-dependent widths, and a variational 
form for the phase: 

\begin{equation}
\chi(\rr,t)=\sum_{\eta=x,y,z} \left[\eta
\alpha_{\eta}(t) + \eta^2 \beta_{\eta}(t)\right] \, ,
\end{equation} 

\noindent where $\alpha_{\eta}(t)$ and $\beta_{\eta}(t)$ are variational 
parameters -- see also \cite{Perez,Yi2}. The linear part of the phase 
ansatz describes the drift of the Fermi fluid (a constant part of the 
velocity field) whereas the quadratic term corresponds to oscillations 
(a linear part of the velocity). This kind of ansatz has proved to 
work very well in reproducing the static properties of trapped dipolar 
fermions \cite{fermionsGoral}. It was also used successfully to model the 
dynamical behavior of trapped dipolar Bose-Einstein condensates 
\cite{Yi2,ExcBosons}. This way, the problem is reduced 
to the study of a few Lagrange equations for the variational parameters 
$w_{\eta}(t)$, $\eta_{0}(t)$, $\alpha_{\eta}(t)$, and $\beta_{\eta}(t)$:
\begin{eqnarray}
\label{var_eqs_start}
&&\ddot{w_{\eta}}=c\frac{\hbar^2}{m^2}\frac{1}{w_{\eta}}
\left[\frac{N}{w_{x}w_{y}w_{z}}\right]^{2/3}
-\omega_{\eta}^{2}w_{\eta}+\frac{2}{Nm}\frac{\partial I}{\partial 
w_{\eta}} \, ,  \\
\label{COM}
&&\ddot{\eta_{0}}+\omega_{\eta}^{2}\eta_{0}=0 \, ,  \\
&&\alpha_{\eta}=\frac{m}{\hbar}\left[\dot{\eta_{0}}
-\frac{\eta_{0}\dot{w_{\eta}}}{w_{\eta}}\right] \, ,  \\      
&&\beta_{\eta}=\frac{m}{2\hbar}\frac{\dot{w_{\eta}}}{w_{\eta}} \, ,
\label{var_eqs_end}
\end{eqnarray}

\noindent where $c=6^{5/3}\sqrt{3/5}\pi^{1/3}/25$ is a numerical factor 
and 
\begin{equation}
I=-\frac{1}{2} \int d\rr \, d\rrp \, n(\rr,t) \, \Vdd(\rr-\rrp) \, 
n(\rrp,t) \, .
\end{equation}

\noindent Again, we can recognize various terms of the Thomas-Fermi theory 
in Eqs.(\ref{var_eqs_start}-\ref{var_eqs_end}): the first term of the 
right-hand side of Eq.(\ref{var_eqs_start}) corresponds to the quantum 
pressure, the second term comes from the trapping potential, and the last 
term results from the dipole-dipole interaction energy.

As we see from Eq.(\ref{COM}), the 
center-of-mass motion (described by $\eta_{0}(t)$) is decoupled and 
follows simple harmonic oscillations, just as in the case of bosons 
\cite{Perez,ExcBosons}. The only independent variables are the widths 
$w_{\eta}(t)$. Their dynamics, governed by Eq.(\ref{var_eqs_start}), can 
be viewed as a motion of a fictitious particle in the following potential:
\begin{equation}
U(w_{x},w_{y},w_{z})=\frac{1}{2}\sum_{\eta}\omega_{\eta}^{2}w_{\eta}^{2}+
c\frac{3\hbar^2}{2m^2}\left[\frac{N}{w_{x}w_{y}w_{z}}\right]^{2/3}-\frac{2I}{Nm}
\end{equation}

\begin{table*}
\begin{tabular}{|c|c|c|c|}
\hline
aspect ratio $\beta$ &
$\beta>0.42$ & \multicolumn{2}{c|}{$\beta \le 0.42$} \\
\hline
lowest mode &
\multicolumn{1}{c|}{quadrupole (unstable for $\beta<5.2$)}&
$N^\sixth\varepsilon$ below the second crossing  &
\multicolumn{1}{c|}{quadrupole} \\ \cline{3-4}
& & $N^\sixth\varepsilon$ above the second crossing  &
\multicolumn{1}{c|}{breathing (unstable)} \\
\hline
\end{tabular}
\caption{Lowest and unstable modes for harmonically trapped dipolar
fermions.}
\label{tabela}
\end{table*}

\begin{figure}[h]
\begin{center}
\includegraphics[width=\columnwidth,clip]{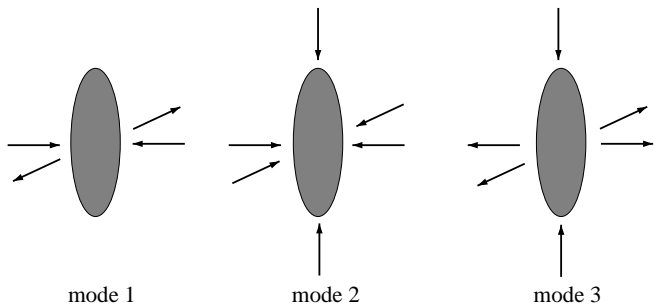}
\caption{
\label{modes}%
Graphical representation of oscillation modes given by the variational 
analysis. We refer to mode 2 as the breathing one and to mode 3 as the 
quadrupole one. Mode 1 will be called the 2D quadrupole mode. }
\end{center}
\end{figure}

By diagonalizing the matrix of second derivatives of this potential 
(evaluated at the stationary point, i.e., for the ground state values of 
the widths -- as calculated in Ref.\cite{fermionsGoral}) one can obtain 
the frequencies and the shapes of three low-lying eigenmodes of the cloud. 
These modes have the same structure as in the bosonic case -- see 
Fig.\ref{modes}.

An interesting question in the analysis of the instability induced by
dipole-dipole interactions concerns the identification of the mode that
becomes unstable (which is indicated by its frequency approaching zero) at
the critical value of the dipolar parameter $N^\sixth\varepsilon$. We have
analyzed this problem for different cylindrically symmetric traps.

\begin{figure}[h]
\begin{center}
\includegraphics[width=\columnwidth,clip]{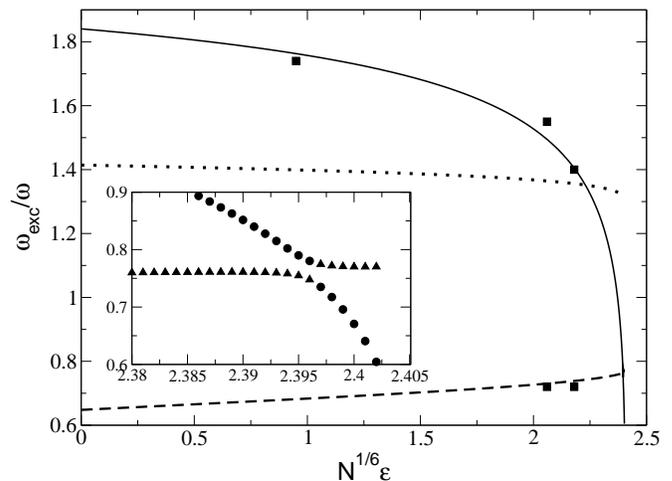}
\caption{
\label{1.54}%
Excitation frequencies of low-lying modes for the dipolar Fermi gas in a
cigar-shaped trap with $\beta=0.42$. The (unstable) mode 2 (breathing) is
depicted with a solid line, while mode 1 (2D quadrupole) with a dotted
line. Mode 3 (quadrupole) is marked by a dashed line. Lines depict the
results of the variational analysis, while squares come from exact
numerical solutions of the hydrodynamic equations. Inset shows a
magnification of the second avoided crossing (mode 3 denoted by
triangles and mode 2 by circles).}
\end{center}
\end{figure}

Let us first concentrate on cigar-shaped traps ($\beta<1$). For 
$0.42<\beta<1$, the lowest mode is always the quadrupole mode (mode 3 in 
Fig.\ref{modes}) and its frequency goes to zero as the critical value of 
$N^\sixth\varepsilon$ is approached. At the characteristic value of 
$\beta \simeq 0.42$ two avoided crossings are present -- see 
Fig.\ref{1.54}. Away from the instability (i.e., for 
$N^\sixth\varepsilon<2.40$) and below both crossings the breathing mode 
(mode 2) has the highest frequency. The quadrupole mode is the lowest one 
and mode 1 (2D quadrupole) is the intermediate one. At the first crossing, 
the nature of two higher modes is exchanged -- now mode 1 (2D quadrupole) 
is the highest one. This first 
crossing is present in all traps with $\beta<1$. At the second crossing 
the breathing mode 2 acquires the lowest frequency and becomes unstable. 
For $N^\sixth\varepsilon$ slightly below the second crossing, the 
oscillation of the quadrupole (lowest) mode is almost purely axial (the 
eigenmode is $(-\delta,+\delta,\sim 1)$, where $\delta$ indicates a very 
small value). Just above the crossing, the shape of the lowest (breathing) 
mode remains very axial with no phase difference between the oscillations 
in the X and Y directions (the eigenmode is now $(+\delta,+\delta,1)$). 
This is always the case for $\beta \le 0.42$. For smaller values of the 
trap aspect ratio, the positions of the two crossings merge (e.g., for 
$\beta=0.031$ the crossings are at $N^\sixth\varepsilon=2.99$, very close 
to the instability). For pancake traps ($\beta>1$), the mode that drives 
the instability (which takes place as long as $\beta<5.2$) is always the 
quadrupole mode. For clarity, the results are summarized in Table 
\ref{tabela}.

The time-dependent variational method employed thus far provides  
approximate solutions of the hydrodynamic equations of the system. To  
confirm their validity, we have performed exact numerical simulations of
the hydrodynamic equations (\ref{hydrodynamic}). First, we find the ground
state of the system by the propagation in imaginary time and use it as an
initial condition for the simulations. Then, we perturb the trapping  
potential for a fraction of a trap period and let the system evolve while
monitoring the resultant oscillations. Finally, we extract the frequency 
of the oscillations and identify the corresponding modes. The results are 
presented in Fig.\ref{1.54} and agree very well with the variational 
solutions.

To summarize, we have investigated the signatures of the mechanical 
instability in a single-component Fermi gas of polarized dipolar 
particles confined in a harmonic trap by studying the spectrum of its 
low-lying hydrodynamic collective excitations. We have employed a 
time-dependent variational method, not previously used in the context of 
excitations of dilute atomic Fermi gases, as well as exact solutions of 
the hydrodynamic equations of the system. We have identified modes 
driving the instability for various geometries of the trapping potential. 
In a rough classification, for oblate and slightly prolate traps 
($\beta>0.42$) it is the quadrupole mode that drives the instability, 
while 
for cigar-shaped traps with $\beta<0.42$ the breathing mode is an 
unstable one. Finally, one should note that the presented analysis 
describes the zero-temperature situation. It would be very interesting, 
although very challenging, to include the effects of temperature in the 
treatment.

K.G. acknowledges support by the Polish KBN grant No. 5 P03B 102 20 and by
the Foundation for Polish Science. Part of the results has been obtained
using computers at Interdisciplinary Centre for Mathematical and
Computational Modelling of Warsaw University.

%\end{multicols}

\end{document}